\documentclass[prl,twocolumn,twoside,preprintnumbers,superscriptaddress,nofootinbib]{revtex4}

\usepackage{amsmath}
\usepackage{graphicx}
\usepackage{dcolumn}
\usepackage{bm}
\usepackage[hyperfootnotes=false]{hyperref}
\usepackage{xspace}

\newcommand{\be}{\begin{equation}}
\newcommand{\ee}{\end{equation}}

\usepackage{color}

\newcommand{\bea}{\begin{eqnarray}}
\newcommand{\eea}{\end{eqnarray}}

\newcommand{ \mysmall}[1]{\scriptscriptstyle #1} % a smaller #

\def\article{\@ifnextchar[{\earticle}{\oarticle}}

\begin{document}
%%%%%%%%%%%%%%%%%%%%%%%%%%%%%%%%%

\title{Contributions of axion-like particles to lepton dipole moments}

\author{W.J.\ Marciano}
\affiliation{Department of Physics, Brookhaven National Laboratory, Upton, NY 11973, USA}
\author{A.\ Masiero}
\affiliation{Dipartimento di Fisica e Astronomia `G. Galilei', Universit\`a di Padova, Italy}
\affiliation{Istituto Nazionale Fisica Nucleare, Sezione di Padova, I--35131 Padova, Italy}
\author{P.\ Paradisi}
\affiliation{Dipartimento di Fisica e Astronomia `G. Galilei', Universit\`a di Padova, Italy}
\affiliation{Istituto Nazionale Fisica Nucleare, Sezione di Padova, I--35131 Padova, Italy}
\author{M.\ Passera}
\affiliation{Istituto Nazionale Fisica Nucleare, Sezione di Padova, I--35131 Padova, Italy}

%\vspace{20pt}

\begin{abstract}
Contributions of a spin 0 axion-like particle (ALP) to lepton dipole moments, $g$-2 and EDMs, are examined. Barr-Zee and light-by-light loop effects from a light pseudoscalar ALP are found to be capable of resolving the long-standing muon $g$-2 discrepancy at the expense of relatively large ALP-$\gamma\gamma$ couplings.  The compatibility of such large couplings with direct experimental constraints and perturbative unitarity bounds is discussed. Future tests of such a scenario are described. 
For CP violating ALP couplings, the electron EDM is found to probe much smaller, theoretically more easily accommodated ALP interactions. Future planned improvement in electron EDM searches is advocated as a way to not
only significantly constrain ALP parameters but also, to potentially unveil a new source of CP violation which could have far reaching ramifications.

\end{abstract}

\maketitle

%%%%%%%%%%%%%%%%%%%%%%%%%%%%%%%%%%%%%%%%%%
\subsection{Introduction}
%%%%%%%%%%%%%%%%%%%%%%%%%%%%%%%%%%%%%%%%%%
Light spin 0 scalars and pseudoscalars, sometimes generically referred to as axion-like-particles (ALPs), often occur in extensions of the Standard Model (SM). Their lightness, relative to the scale of new physics (NP) from which they stem, can be understood in terms of their pseudo-Goldstone boson nature, i.e.\ connection with an underlying broken symmetry. ALPs are a generalization of the well-known QCD axion, but with the caveat that their mass and couplings to other particles are arbitrary parameters to be determined or bounded by experiment. In that context, we concentrate here on ALP couplings to photons (ALP-$\gamma\gamma$ interactions)~\cite{Jaeckel:2010ni} and their Yukawa couplings to leptons. We restrict our attention to ALPs in the approximate mass range of $100\,{\rm MeV}$--$1\,{\rm GeV}$ where experimental constraints~\cite{Jaeckel:2010ni,Jaeckel:2015jla} are currently rather loose, leaving open the possibility of potentially new observable effects.

In this study, we examine indirect effects of ALPs on lepton electromagnetic dipole moments. For the mass range and couplings considered, the muon anomalous magnetic moment $a_\mu = (g-2)_\mu /2$ provides a potentially sensitive probe of NP~\cite{gm2_rev_NP1,Giudice:2012ms}. Currently, comparison of the SM prediction with the experimental value shows an interesting $\sim 3.4 \,\sigma$ discrepancy, 
\be
\Delta a_\mu = a_\mu^{\mysmall \rm EXP}-a_\mu^{\mysmall \rm SM} = 273 \, (80) \times 10^{-11}\,,
\label{eq:gmu}
\ee
based on $a_\mu^{\mysmall \rm EXP}=116592091(63)\times10^{-11}$~\cite{bnl} and $a_\mu^{\mysmall \rm SM}=116591818(49)\times 10^{-11}$~\cite{Knecht:2001qg, Knecht:2001qf, Blokland:2001pb, RamseyMusolf:2002cy, Melnikov:2006sr, Prades:2009tw, Aoyama:2012wk, EW, Jegerlehner:2009ry, HLMNT11, DHMZ11, HadronicNLO}. For an alternative up-to-date analysis that leads to a larger $4.0 \,\sigma$ discrepancy see ref.~\cite{Jegerlehner:2015stw}. On the theory side, there is a fairly general consensus that hadronic loop uncertainties alone cannot explain such a large discrepancy. Nevertheless, considerable effort is being expended to reduce the uncertainty in the SM prediction. Regarding the experimental result~\cite{bnl} in eq.~(\ref{eq:gmu}), an anticipated new measurement at Fermilab, E989, is expected to improve the precision by a factor of four~\cite{Grange:2015fou}. In addition, a completely new low-energy approach to measuring the muon $g$-2 is being developed by the E34 collaboration at J-PARC~\cite{J-PARC}. In a few years, we should know much better whether the discrepancy in eq.~(\ref{eq:gmu}) is due to NP. For comparison, we note that for the electron 
$\Delta a_e = -91(82)\times10^{-14}$~\cite{Aoyama:2014sxa}, i.e.\ relatively good agreement between theory and experiment.

\begin{figure}
\includegraphics[width=0.9\linewidth]{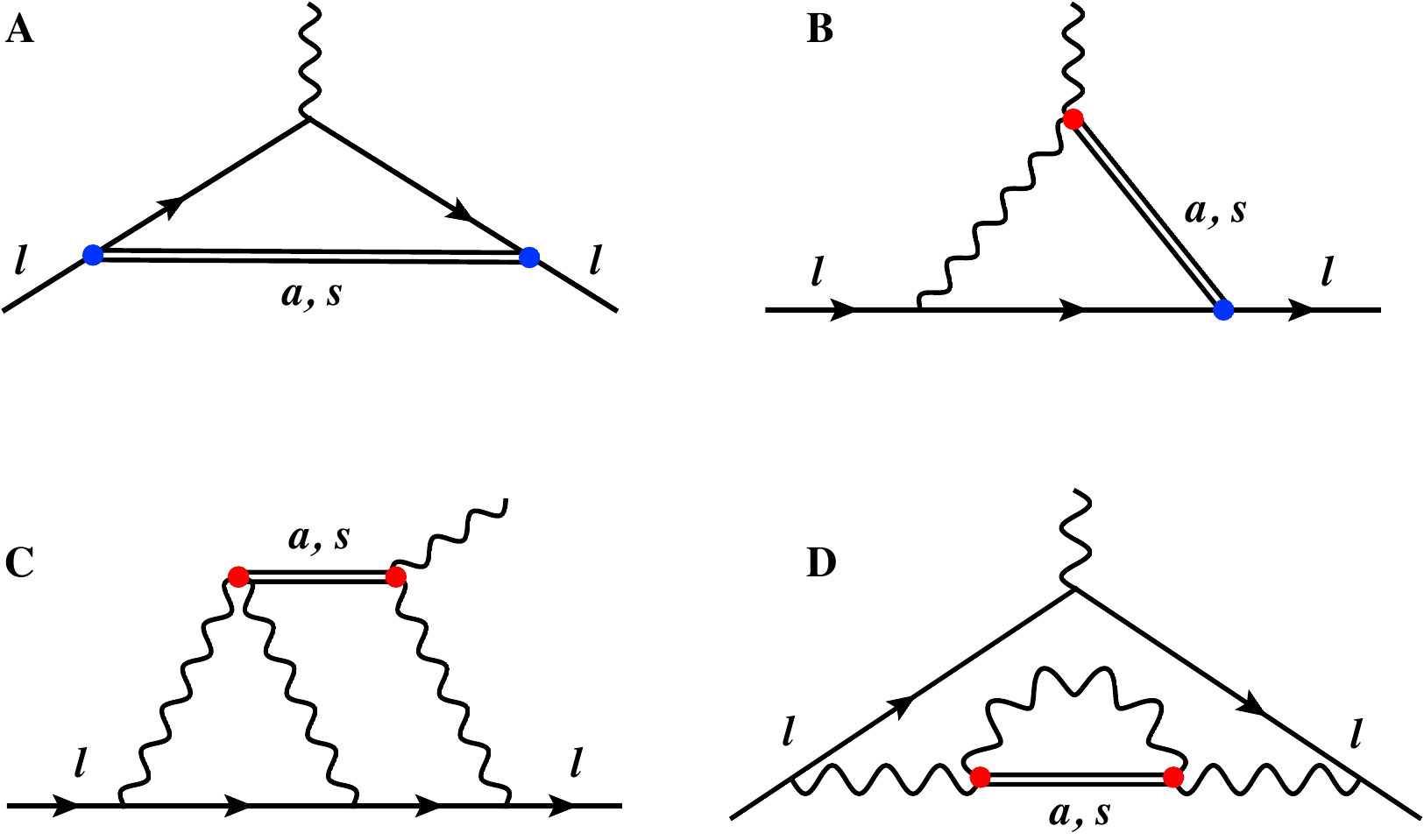}
%\vspace{-20pt}
\caption{
Representative contributions of a scalar `$s$' and a pseudoscalar `$a$' ALP to the lepton `$\ell$' dipole moments.
}
\label{fig:LbL}
\end{figure}

Already, a possible resolution of the muon $g$-2 discrepancy by one-loop contributions from scalar particles with relatively large Yukawa couplings to muons, of $\mathcal{O}(10^{-3})$, 
has been considered~\cite{Chen:2015vqy} (see fig.~\ref{fig:LbL}A). In the case of a pseudoscalar, the one-loop contribution had the wrong sign to resolve the discrepancy on its own.
Here, we extend that discussion to include ALP-$\gamma\gamma$ couplings as well as Yukawa couplings. In that way, two new ALP contributions to lepton dipole moments are potentially 
important:  
i) Barr-Zee (BZ)~\cite{Barr:1990vd} one-loop diagrams that involve both ALP-$\gamma\gamma$ and ALP Yukawa interactions with leptons (see fig.~\ref{fig:LbL}B)
and 
ii) two-loop light-by-light (LbL) and vacuum polarisation diagrams stemming only from ALP-$\gamma \gamma$ interactions (see fig.~\ref{fig:LbL}C, \ref{fig:LbL}D).
As we shall show, for relatively large ALP-$\gamma\gamma$ couplings, they can potentially resolve (fully or partially) the muon $g$-2 discrepancy. In fact, even with a fairly large negative pseudoscalar contribution from fig.~\ref{fig:LbL}A, their positive contribution can dominate.

If ALPs have both CP even and odd components, their combination can lead to CP violating fermion electric dipole moments (EDMs) through the diagrams in fig.~\ref{fig:LbL}.
One-loop pure Yukawa diagrams (see fig.~\ref{fig:LbL}A) have been already considered in~\cite{Chen:2015vqy}. 
In this case, the electron EDM $d_e$, which is currently constrained at the level of $|d_e| \leq 8.7 \times 10^{-29}e~$cm~\cite{Baron:2013eja}, turns out to be a very sensitive probe of our scenario. The expected future experimental sensitivity $|d_e| \lesssim 10^{-30}e~$cm~\cite{Hewett:2012ns} will further strengthen the impact of this observable and nicely complement the SHiP proposal~\cite{Alekhin:2015byh} at CERN's SPS fixed target facility. That experiment is intended (among its many goals) to directly search for ALPs produced via the Primakoff effect in a dense target. 

The scenario we are advancing, requires relatively large ALP-$\gamma\gamma$ couplings. For that reason, we will address current and potential future direct experimental constraints on such a coupling as well as a possible breakdown of perturbative unitarity in the diagrammatic use of such effective couplings in loop calculations.

%
%%%%%%%%%%%%%%%%%%%%%%%%%%%%%%%%%%%%%%%%%%
%%%%%%%%%%%%%%%%%%%%%%%%%%%%%%%%%%%%%%%%%%

%%%%%%%%%%%%%%%%%%%%%%%%%%%%%%%%%%%%%%%%%%
\subsection{ALPs contributions to lepton \boldmath $g$-2 \unboldmath}
%%%%%%%%%%%%%%%%%%%%%%%%%%%%%%%%%%%%%%%%%%
The possibility that the discrepancy in eq.~(\ref{eq:gmu}) is a NP signal 
has been widely discussed in the literature for a number of NP scenarios~\cite{gm2_rev_NP1,Giudice:2012ms}.
Here, we examine contributions to the lepton $g$-2 induced by ALPs primarily coupled to photons and leptons.
In general, ALPs can be scalars or pseudoscalars (or mixed if CP is violated). 
The effective Lagrangian (assumed valid for scales $< \mathcal{O}{(\rm TeV)}$) describing the interactions of a pseudoscalar ALP `$a$' 
with photons and SM fermions $\psi$ can be parametrized by:
\be
\mathcal{L} = \frac{1}{4}\, g_{a\gamma\gamma} \, a \, F_{\mu\nu} {\tilde F}^{\mu\nu} 
+ i \, y_{a\psi}\, a\,\bar{\psi}\gamma_5 \psi \,,
\label{eq:WZW}
\ee
where $g_{a\gamma\gamma}$ is a dimensionful coupling, $F_{\mu\nu}$ and ${\tilde F}^{\mu\nu}$ are the electromagnetic tensor and its dual, respectively, 
and $y_{a\psi}$ are real dimensionless Yukawa couplings. 
The first term of eq.~(\ref{eq:WZW}) reproduces the well-known $\pi^0 \to \gamma\gamma$ anomalous coupling 
for $a \equiv \pi^0$ and $g_{\pi^0\gamma\gamma}\equiv \alpha/ (\pi F_\pi)$, where $\alpha$ is the fine-structure constant and $F_\pi = 92$ MeV is the neutral pion decay constant. 
For the scalar case, replace ${\tilde F}^{\mu\nu}$ by $F_{\mu\nu}$, `$a$' by `$s$', and delete $i\gamma_5$.

In the SM, the UV cut-off of the effective theory can be roughly interpreted as the cut-off scale of chiral perturbation theory 
$2\sqrt{2}\pi F_\pi = 820~$MeV or approximately $m_\rho \sim 770~$MeV.  Therefore, a natural parametrization of $g_{a\gamma\gamma}$ is:
\be
g_{a\gamma\gamma} \equiv \frac{2\sqrt{2}\,\alpha}{\Lambda}\, c_{a\gamma\gamma}\,,
\label{eq:coefficients}
\ee
where $\Lambda$ is the NP UV cut-off while $c_{a\gamma\gamma}$ is a dimensionless coupling. In the case
of the pion, $c_{\pi^0\gamma\gamma} \sim 1$. 

The effective $a\gamma\gamma$ and $y_{a\ell}$ vertices induce contributions to the $g$-2 of a lepton $\ell$ via 
one-loop BZ diagrams and two-loop LbL diagrams (analogous to the SM hadronic LbL contribution of the neutral pion 
exchange~\cite{Knecht:2001qg, Knecht:2001qf, Blokland:2001pb, RamseyMusolf:2002cy, Melnikov:2006sr, Prades:2009tw}) 
shown in fig.~\ref{fig:LbL}.
In particular, by an explicit calculation, we find the following effects (assuming the point-like couplings of eq.~(\ref{eq:WZW}) and a sharp cut-off $\Lambda$):
\begin{align}
a_{\ell,a}^{{\rm \mysmall BZ}} &\simeq \left ( \frac{m_\ell}{4\pi^2} \right ) 
\,g_{a\gamma\gamma}\, y_{a\ell} \,\ln \frac{\Lambda }{m_a} \,,
\label{eq:gm2_BZ}
\\
a_{\ell,a}^{{\rm \mysmall LbL}}
&\simeq \, 3  \, \frac{\alpha}{\pi}  
\left(\frac{m_\ell \, g_{a\gamma\gamma}}{4\pi}\right)^2
\ln^{2} \! \frac{\Lambda }{m_a}\,,
\label{eq:gm2_LbL}
\end{align}
where $m_a$ is the ALP's mass and we kept only the leading log-enhanced terms since they should provide the main ALP contribution 
to the lepton $g$-2 for $\Lambda \sim 1$~TeV and $m_a \lesssim 1$~GeV.
In deriving eqs.~(\ref{eq:gm2_BZ},\ref{eq:gm2_LbL}) as well as subsequent loop effects, we assume that $g_{a\gamma\gamma}$ remains essentially 
constant throughout the integration over virtual photon-loop momentum $0 < |k^2| < \Lambda^2$.  That requires an effective point-like coupling 
$g_{a\gamma\gamma}$ arising from high-mass scale phenomena of $\mathcal{O}{(\Lambda)}$.  

A muon $g$-2 realization of our generic Barr-Zee analysis for a (pseudo)scalar with a relatively large $\gamma\gamma$ coupling
induced by heavy fermion triangle diagrams has been considered some time ago in~ \cite{Chang:2000ii,Cheung:2001hz}.
We note that for a single fermion triangle diagram with the same magnitude Yukawa couplings for a pseudoscalar and scalar, the effective loop-induced $g_{a\gamma\gamma}$ coupling is a factor of $-3/2$ times the effective $g_{s\gamma\gamma}$ coupling. This factor tends to make $\gamma\gamma$ or $gg$ production of a pseudoscalar more likely than a scalar.

An inspection of eqs.~(\ref{eq:gm2_BZ},\ref{eq:gm2_LbL}) leads to the following remarks: 
\begin{itemize}
\item The sign of $a_{\ell,a}^{{\rm \mysmall BZ}}$ depends on the sign of the product $g_{a\gamma\gamma}\, y_{a\ell}$ while that of  
$a_{\ell,a}^{{\rm \mysmall LbL}}$ is positive, as needed to accommodate the $\Delta a_\mu$ discrepancy (see eq.~(\ref{eq:gmu})). 
In the case of a scalar ALP, the leading LbL contribution changes sign~\cite{Blokland:2001pb}, 
while, for our convention, the BZ doesn't. If LBL is taken in isolation, that would imply the indirect bound $g_{s\gamma\gamma} < \mathcal{O}(10^{-3}{\rm GeV}^{-1})$.
\item $a_{\ell,a}^{{\rm \mysmall LbL}}$ follows the expected scaling $a_{\ell,a}^{{\rm \mysmall LbL}} \!\propto\! m^2_{\ell}$ (this is also true for $a_{\ell,a}^{{\rm \mysmall BZ}}$ if 
$y_{a\mu}/y_{ae} \!\sim\! m_\mu/m_e$). Combining the NP sensitivity and the present experimental resolutions on the lepton $g$-2, it turns out that $a_{\mu}$ rather than $a_e$ is 
the better probe of our NP scenario. 
\item The BZ contribution accommodates the muon $g$-2 discrepancy for $g_{a\gamma\gamma} y_{a\mu} \approx 10^{-7}\,{\rm GeV^{-1}}$. Its effect is typically larger than 
the LbL unless $y_{a\mu}$ is very small. 
\end{itemize}

\begin{figure}
\includegraphics[width=0.9\linewidth]{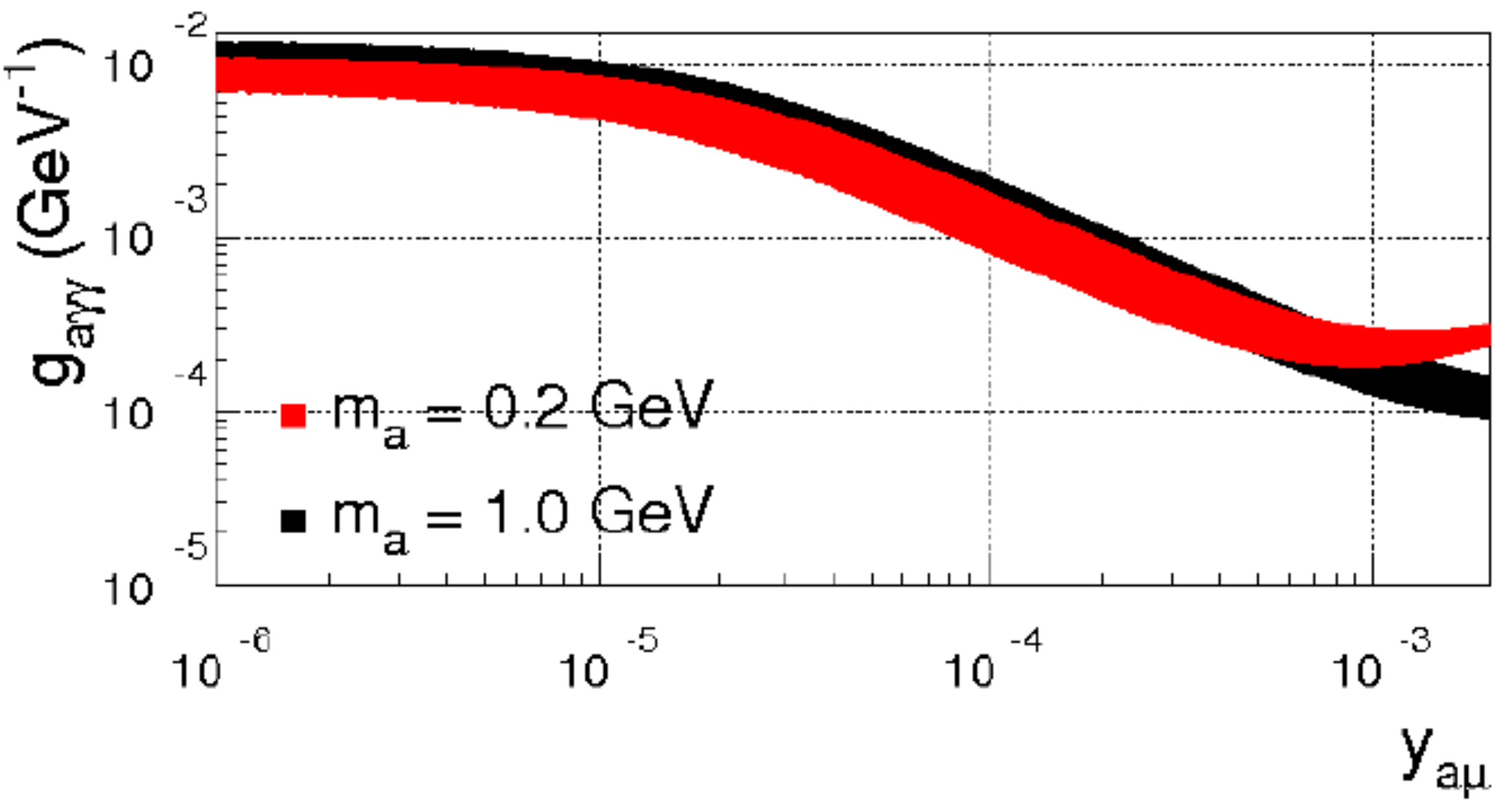}
\includegraphics[width=0.9\linewidth]{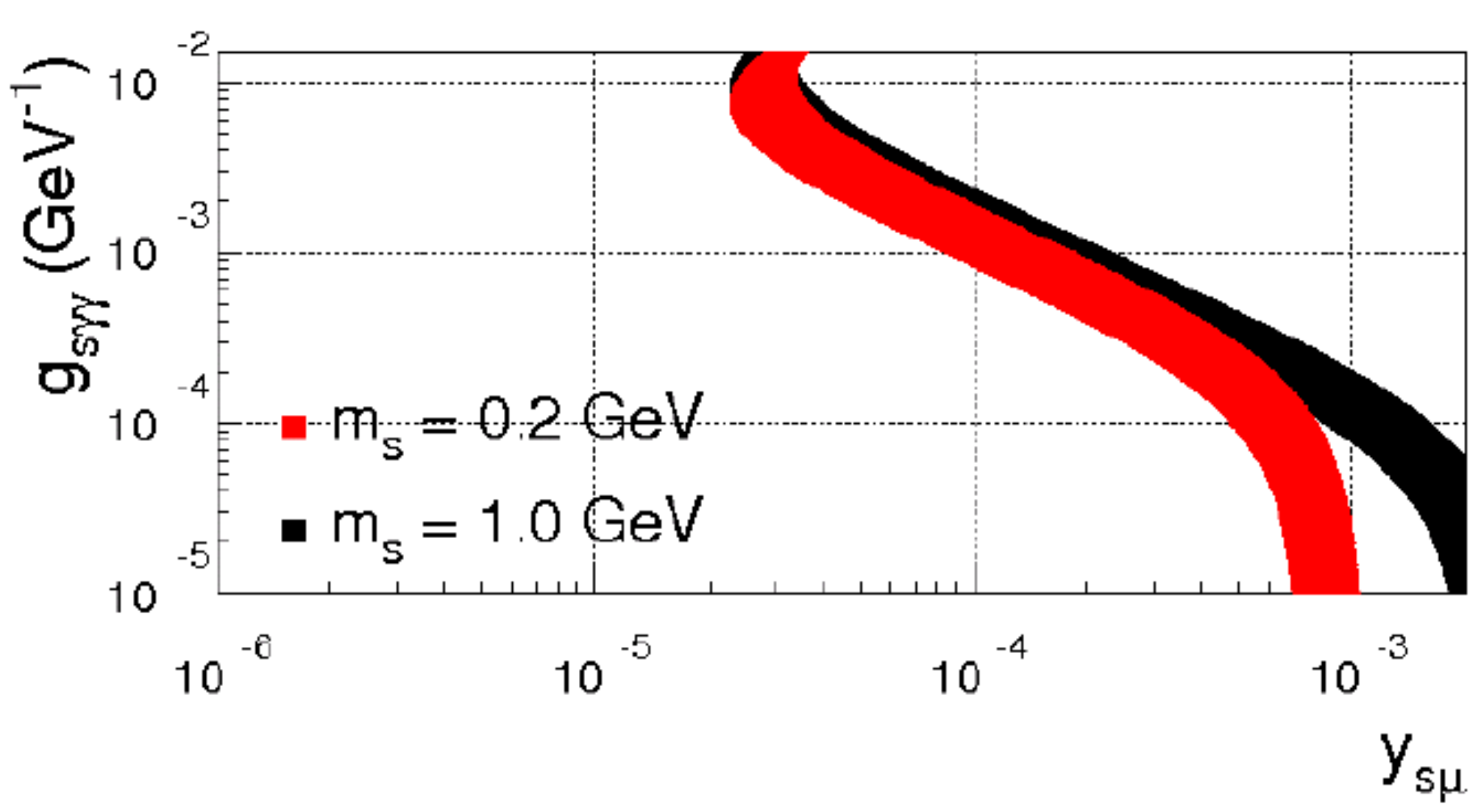}
\vspace{-10pt}
\caption{
Pseudoscalar (upper) and scalar (lower) $1\sigma$ solution bands to the $\Delta a_\mu$ 
discrepancy as a function of $y_{a\mu}$ and $g_{a\gamma\gamma}$ for the pseudoscalar 
and $y_{s\mu}$ and $g_{s\gamma\gamma}$ for the scalar.  They correspond to the sum of 
pure Yukawa~\cite{Chen:2015vqy}, BZ and LbL contributions with BZ taken to be positive 
and $\Lambda = 1$ TeV.
We have truncated the bands at $|y_{a\mu}|, |y_{s\mu}| < 2\times10^{-3}$ in order to avoid
experimental constraints~\cite{Batell:2016ove}.
}
\label{fig:ma_cagg_plane}
\end{figure}

In fig.~\ref{fig:ma_cagg_plane}, we illustrate by $1\sigma$ bands, pseudoscalar (upper) and scalar (lower) ALP solutions to the muon $g$-2 discrepancy as a function 
of $y_{a\mu}$ and $g_{a\gamma\gamma}$. They correspond to the sum of the pure one-loop Yukawa contribution given in ref.~\cite{Chen:2015vqy} along with 
BZ (eq.~(\ref{eq:gm2_BZ})) and LbL (eq.~(\ref{eq:gm2_LbL})) contributions for $\Lambda= 1$ TeV. 
For a scalar ALP, the BZ depends on the sign of $y_{s\mu} \, g_{s\gamma\gamma}$ and LbL changes sign.  
Note that, although the pure one-loop Yukawa contribution is negative for a pseudoscalar ALP~\cite{Chen:2015vqy}, BZ and LbL (for positive $y_{a\mu} \, g_{a\gamma\gamma}$) 
dominate the solution, solving the muon $g$-2 discrepancy for $10^{-4} \lesssim g_{a\gamma\gamma}({\rm GeV}^{-1}) \lesssim10^{-2}$. 
That corresponds, from eq.~(\ref{eq:coefficients}) with $\Lambda = 1$ TeV, to a $c_{a\gamma\gamma}$ in the range $5 - 500$.
We do not attempt to construct a realistic model with such large ALP coupling to photons, but note that it is likely to require a new type of non-perturbative dynamics and/or a 
high multiplicity of heavy states contributing to $g_{a\gamma\gamma}$ at the loop level. For a scalar ALP, there are solutions to the muon $g$-2 discrepancy (see lower 
fig.~\ref{fig:ma_cagg_plane}) dominated by pure one-loop Yukawa contributions for $y_{s\mu} \sim \mathcal{O}(10^{-3})$ and $g_{s\gamma\gamma} < 10^{-4}\,{\rm GeV}^{-1}$.
Note, for both plots we assume the rather conservative bound $|y_{a\mu}|, |y_{s\mu}| < 2\times10^{-3}$, in keeping with BABAR studies of 
$e^+e^- \to \mu^+\mu^-\mu^+\mu^-$~\cite{Batell:2016ove}.

%
%%%%%%%%%%%%%%%%%%%%%%%%%%%%%%%%%%%%%%%%%%
%%%%%%%%%%%%%%%%%%%%%%%%%%%%%%%%%%%%%%%%%%

The above remarks raise the question whether perturbative unitarity is respected for such a large $g_{a\gamma\gamma}$ coupling. We therefore computed the partial wave unitarity bounds of 
$\gamma\gamma \to \gamma\gamma$ amplitudes mediated by a pseudoscalar `$a$', obtaining
\begin{equation}
\sqrt{s} < 4\sqrt{2\pi}\,\, g^{-1}_{a\gamma\gamma}.
\label{eq:bound_unitary}
\end{equation}
For $g_{a\gamma\gamma} = 10^{-2} \,{\rm GeV^{-1}}$, eq.~(\ref{eq:bound_unitary}) gives $ \sqrt{s} \,\lesssim 1 \,{\rm TeV}$
while for $g_{a\gamma\gamma} = 10^{-4} \, {\rm GeV^{-1}}$ it implies $ \sqrt{s} \,\lesssim 100 \,{\rm TeV}$. 
The calculation of the partial wave unitary constraint from the process 
$a\gamma \to a\gamma$ leads to the same result. Therefore, our effective theory remains unitary up to energies at or above the TeV scale; 
at even higher scales NP unitarization is expected. If other scattering channels exist with larger couplings, e.g.\ $g g \to g g$, perturbative 
unitarity may break down well before the TeV scale. 

The effective coupling $a\gamma\gamma$ in our Lagrangian also induces a photon vacuum polarization which provides another contribution to the lepton $g$-2 
(see fig.~\ref{fig:LbL}D).
In this case there is an analogous SM effect, arising from the $\pi^0$ exchange, which is included in the hadronic contribution to $a_\mu^{\mysmall \rm SM}$ 
through the dispersive calculation~\cite{Jegerlehner:2009ry,HLMNT11,DHMZ11,Achasov:2002bh}. If we keep only the dominant log-enhanced term, we find
\be
a_{\ell,a}^{{\rm \mysmall VP}} \, \simeq \,  \frac{\alpha}{\pi} \left(\frac{m_\ell \, g_{a\gamma\gamma}}{12 \pi}\right)^2 \ln \frac{\Lambda}{m_a}.
\label{eq:gm2_vp}
\ee
For a scalar ALP, just replace `$a$' by `$s$' in eq.~(\ref{eq:gm2_vp}).
However, employing $g_{a\gamma\gamma} \lesssim 10^{-2}\,\rm{GeV}^{-1}$~\cite{Jaeckel:2015jla} in eq.~(\ref{eq:gm2_vp}) we obtain
$a_{\mu,a}^{{\rm \mysmall VP}} \lesssim  2\times 10^{-11}$ which is much smaller than the LbL effect in eq.~(\ref{eq:gm2_LbL}) 
and can therefore be neglected.

%%%%%%%%%%%%%%%%%%%%%%%%%%%%%%%%%%%%%%%%%%
\subsection{ALPs contributions to lepton EDMs}
%%%%%%%%%%%%%%%%%%%%%%%%%%%%%%%%%%%%%%%%%%
So far we restricted our discussion to the effects induced by pure pseudoscalar or scalar bosons to the lepton $g$-2 since the Lagrangian in eq.~(\ref{eq:WZW}) is CP conserving. 
However, more generally, if the scalar and pseudoscalar states mix due to the presence of CP violating sources, lepton EDMs $d_\ell$ are also generated. Calling $\Phi$ this 
mixed state, we can generalise the Lagrangian of eq.~(\ref{eq:WZW}) as follows,
\be
\mathcal{L} = 
\frac{\tilde g_{\Phi\gamma\gamma}}{4} \Phi F {\tilde F} + \frac{g_{\Phi\gamma\gamma}}{4} \Phi F^2 + 
\left( y_{\Phi\psi} \Phi\bar{\psi} P_L \psi + h.c. \right),
\label{eq:WZW2}
\ee
where $y_{\Phi\psi}$ is a complex Yukawa coupling and $P_L = (1-\gamma_5)/2$.  
Starting from the above Lagrangian, we can compute the leading BZ contributions to $a_\ell$ and $d_\ell$:
\begin{align}
&
\!\!\!a_{\ell}^{{\rm \mysmall BZ}} \simeq 
m_\ell 
\!\left [ 
\frac{g_{\Phi\gamma\gamma}\, {\rm Re}(y_{\Phi\ell}) \!+ \tilde g_{\Phi\gamma\gamma}\, {\rm Im}(y_{\Phi\ell}) }{4\pi^2} 
\right ] \ln \frac{\Lambda }{m_\Phi},
\label{eq:dipoles_BZ1}
\\
&
\!\!\!
\frac{d_{\ell,\Phi}^{{\rm \mysmall BZ}}}{e} 
\simeq 
\frac{g_{\Phi\gamma\gamma}\,{\rm Im}(y_{\Phi \ell}) + \tilde g_{\Phi\gamma\gamma}\,{\rm Re}(y_{\Phi \ell}) }{8\pi^2} \, \ln \frac{\Lambda }{m_\Phi},
\label{eq:dipoles_BZ2}
\end{align}
as well as the corresponding LbL contributions:
\begin{align}
a_{\ell,\Phi}^{{\rm \mysmall LbL}} & \, \simeq \, 
3 \, \frac{\alpha}{\pi}  \,
\frac{m^2_\ell}{16\pi^2}
\left(\tilde g^2_{\Phi\gamma\gamma} - g^2_{\Phi\gamma\gamma} \right)
\ln^{2}\! \frac{\Lambda }{m_\Phi},
\\
\frac{d_{\ell,\Phi}^{{\rm \mysmall LbL}}}{e} & \, \simeq \, 
3 \, \frac{\alpha}{\pi}  \,
\frac{m_\ell}{16\pi^2}
\left(g_{\Phi\gamma\gamma} \,\tilde g_{\Phi\gamma\gamma} \right) \,
\ln^{2}\! \frac{\Lambda }{m_\Phi}.
\label{eq:dipoles}
\end{align}
Neglecting one-loop pure Yukawa diagrams already considered in~\cite{Chen:2015vqy}, the experimental bound~\cite{Baron:2013eja} on $d_e$ is satisfied for 
\begin{align} 
|g_{\Phi\gamma\gamma}\,{\rm Im}(y_{\Phi e})|, |\tilde g_{\Phi\gamma\gamma}\,{\rm Re}(y_{\Phi e})| &\lesssim 5\!\times\! 10^{-14} \,{\rm GeV}^{-1}\,,
\label{eq:bound_edm_BZ}
\\
\sqrt{|g_{\Phi\gamma\gamma} \,\,\tilde g_{\Phi\gamma\gamma}|}
&\lesssim 6\!\times\! 10^{-5}\, {\rm GeV}^{-1}\,,
\label{eq:bound_edm}
\end{align}
where we assumed masses for $\Phi$ in the range $ 0.1 \lesssim m_\Phi (\rm{GeV}) \lesssim 1$ and $\Lambda =1$ TeV. For CP violating phases of $\mathcal{O}(1)$, 
that is $g_{\Phi\gamma\gamma} \sim \tilde g_{\Phi\gamma\gamma}$, LbL effects to $d_e$ are already probing the TeV scale provided $c_{a\gamma\gamma}\sim \mathcal{O}(1)$, 
see eq.~(\ref{eq:coefficients}). A sensitivity up to a scale of $\Lambda\sim 10$ TeV could be reached in the future thanks to the expected experimental sensitivity 
$|d_e| \lesssim 10^{-30}e~$cm~\cite{Hewett:2012ns}. Such high-scale ALP interactions could also be studied by the SHiP proposal~\cite{Alekhin:2015byh}.
The BZ contribution to $d_e$ is much larger than the LbL one unless $y_{\Phi e}$ is very small, as shown by eqs.~(\ref{eq:bound_edm_BZ}, \ref{eq:bound_edm}).

%%%%%%%%%%%%%%%%%%%%%%%%%%%%%%%%%%%%%%%%%%%%%%%%
\subsection{Experimental tests at \boldmath $e^+e^-$ \unboldmath colliders}
%%%%%%%%%%%%%%%%%%%%%%%%%%%%%%%%%%%%%%%%%%%%%%%%

As recently shown in~\cite{Jaeckel:2015jla,Knapen:2016moh}, $e^+e^-$ colliders can set bounds on ALP-$\gamma\gamma$ 
couplings over a broad range of ALP masses. In particular, the pseudoscalar ALP production mechanism proceeds through the 
process $e^+e^- \to \gamma^* \to\gamma a$ which is characterised by the following differential cross-section:
\begin{equation}
\left(\frac{d\sigma}{d\cos\theta}\right)_{\gamma a}  = 
\frac{\alpha}{64} \,
g^2_{a\gamma\gamma} \left(1 - \frac{m^2_a}{s}\right)^{3} (1+\cos^2\theta)\,,
\label{eq:cross-section_LEP}
\end{equation}
where $\theta$ is the angle between the ALP and the beam axis in the center-of-mass.
For $m_a \lesssim 1\,{\rm GeV}$, the process $e^+e^- \to \gamma a$ at very high energies (e.g., LEPII) followed by $a\to\gamma\gamma$ could simulate 
the process $e^+e^-\to 2\gamma$, since the two photons from $a\to\gamma\gamma$ are very collimated. 
With this assumption, the authors of ref.~\cite{Knapen:2016moh} suggest a bound $g_{a\gamma\gamma} \lesssim 10^{-3}~{\rm GeV}^{-1}$ based on 
$e^+e^-\to 2\gamma$ OPAL data~\cite{Abbiendi:2002je}, but no detailed discussion is given. Although we agree that LEPII data can likely provide a better 
constraint than the $g_{a\gamma\gamma} \lesssim 10^{-2}~{\rm GeV}^{-1}$ bound of ref.~\cite{Jaeckel:2015jla}, a detailed study of detector acceptances 
and efficiencies is required before drawing firm conclusions~\cite{Abbiendi_Raggi}. Here, we note that the more restrictive 
$\mathcal{O}(10^{-3}{\rm GeV}^{-1})$ bound on $g_{a\gamma\gamma}$ would significantly reduce the LbL contribution, while the BZ one could still provide 
a solution to the muon $g$-2 discrepancy for $y_{a\mu}\gtrsim 10^{-4}$, see upper fig.~\ref{fig:ma_cagg_plane}. 

\begin{figure}
\includegraphics[width=0.9\linewidth]{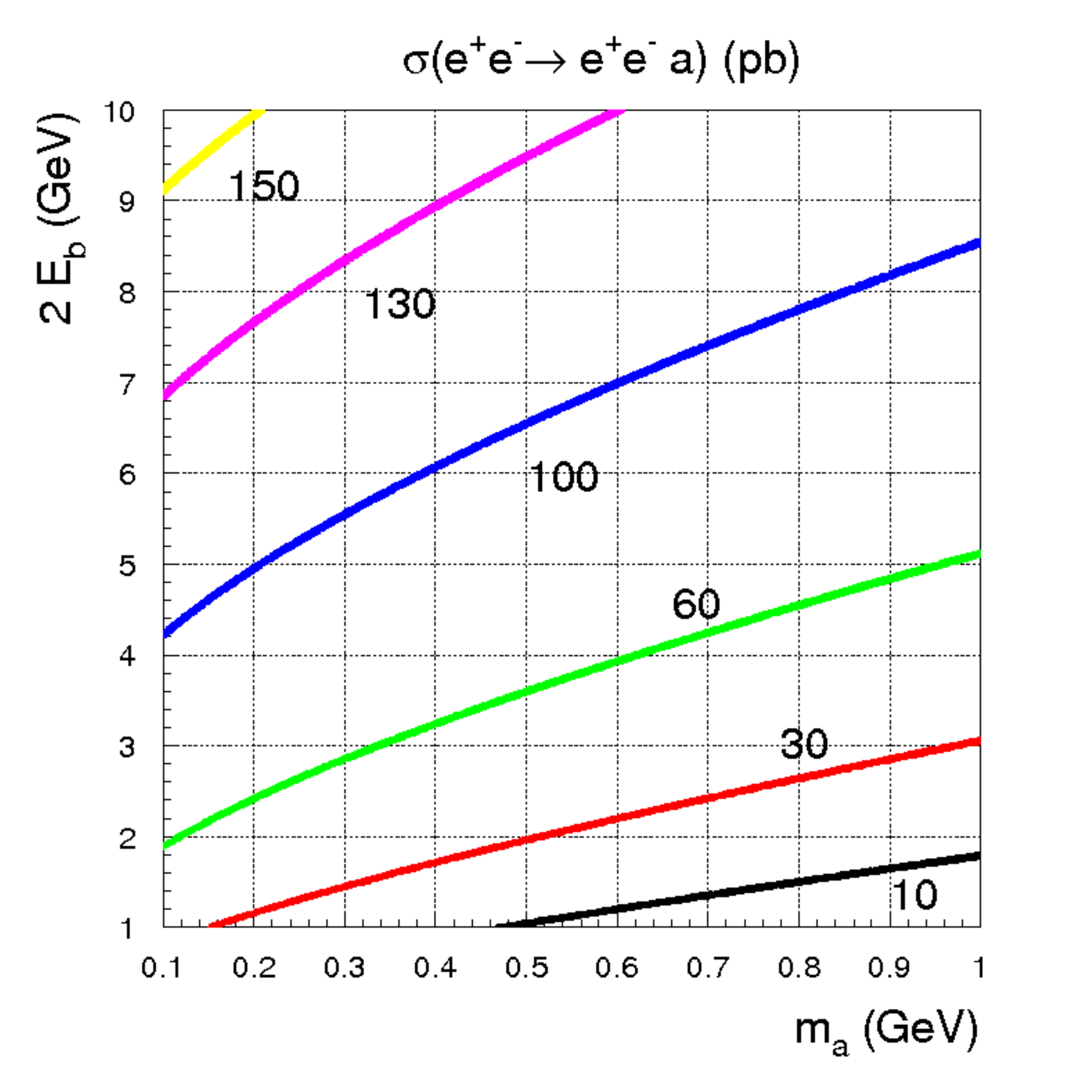}
\caption{Contour plot for $\sigma(e^+e^-\to e^+e^-a)$ in the ($m_a, 2E_b$) plane imposing $g_{a\gamma\gamma} = 10^{-2}\,{\rm GeV}^{-1}$.
For smaller $g_{a\gamma\gamma}$ the cross-section is quadratically reduced.}
\label{fig:sigma_gm2}
\end{figure}

In the following, we focus on direct experimental searches for ALPs with dominant $\gamma \gamma$ couplings and masses up to a few GeV 
at low-energy $e^+e^-$ colliders. The relevant processes are
\begin{eqnarray}
e^+e^- & \to & e^+e^- \gamma^\ast \gamma^\ast \to e^+e^- a\,,
\label{eq:tchannel}
\\
e^+e^- & \to & \gamma^\ast \to \gamma a\,,
\label{eq:schannel}
\end{eqnarray}
where the production cross--section $\sigma(e^+e^- \to e^+e^- a)\equiv \sigma_{eea}$ is dominated by the $t$-channel with quasi-real photons, 
especially for $\sqrt{s} \gtrsim 1~$GeV. In the equivalent photon approximation, the total cross-section $\sigma_{eea}$ reads~\cite{Brodsky:1971ud}
\begin{equation}
\sigma_{eea}  \simeq \frac{\alpha^2}{4\pi} \,\, g^{2}_{a\gamma\gamma}
\left(\ln\frac{E_b}{m_e}\right)^{\!2} \!\! f \!\left(\frac{m_a}{2E_b}\right)\;, 
\label{eq:sigma}
\end{equation}
where $E_b\equiv \sqrt{s}/2$ is the beam energy and $f(z)$ is
\begin{equation}
f(z) = (z^2-1) (z^2+3) - (z^2+2)^2 \ln z \,.
\label{eq:fz}
\end{equation}
If we take, for example, $\sqrt{s}=1$ GeV and $m_a=m_{\pi^0}$, we find
\begin{align}
\sigma_{eea}(\sqrt{s}=1\,{\rm GeV}) &\approx 31 \, {\rm pb} \, \left(\frac{g_{a\gamma\gamma}}{10^{-2}\,{\rm GeV}^{-1}}\right)^{\!2}\,, 
\\
\sigma_{\gamma a}(\sqrt{s} =1\,{\rm GeV}) &\approx 9 \, {\rm pb} \, \left(\frac{g_{a\gamma\gamma}}{10^{-2}\,{\rm GeV}^{-1}}\right)^{\!2}\,, 
\label{eq:gm2_sigma_num}
\end{align}
where $\sigma_{\gamma a}$ is the total cross-section obtained by integrating  $\left(\frac{d\sigma}{d\cos\theta}\right)_{\gamma a}$ over $\cos\theta$.
Let us focus on the process in (\ref{eq:tchannel}), as it is the most sensitive to NP effects for $\sqrt{s}\gtrsim 1\,$GeV. Figure~\ref{fig:sigma_gm2} 
shows the predictions for $\sigma_{eea}$, in the plane ($m_a, 2E_b$),  imposing $g_{a\gamma\gamma} = 10^{-2}\,{\rm GeV}^{-1}$.
We consider beam energies in the range 1 $\leq \sqrt{s}\,({\rm GeV})\leq 10$ in order to monitor the signal cross-section that could be expected at the KLOE2~\cite{kloe2}, 
CMD3~\cite{cmd3}, SND~\cite{snd}, BES III~\cite{bes3} and Belle II~\cite{belle2} experiments. 
For comparison, the SM production cross-sections for $a \equiv \pi^0, \eta$ computed using eq.~(\ref{eq:sigma}) at $\sqrt{s}=1,2,10$~GeV are $\sigma_{ee\pi} \simeq 0.3, 0.5, 1.2$~nb and 
$\sigma_{ee\eta} \simeq 0.04, 0.2, 0.8$~nb, respectively. 
As illustrated by fig.~\ref{fig:sigma_gm2}, $\sigma_{eea}$ grows with energy; whereas $\sigma_{\gamma a}$ is essentially constant even well above threshold.
Although we are not aware of any dedicated search for  the non-standard process $e^+e^-\to e^+e^- a$, we believe it would be worthwhile studying it at running 
and upcoming $e^+e^-$ colliders.
%
%%%%%%%%%%%%%%%%%%%%%%%%%%%%%%%%%%%%%%%%%%
%%%%%%%%%%%%%%%%%%%%%%%%%%%%%%%%%%%%%%%%%%

ALPs decay widths could perhaps be measured at JLab via the Primakoff effect. Indeed, with the advent of JLab's 12~GeV upgrade, which aims at gathering high precision measurements 
of the two-photon decay widths of $\eta$ and $\eta^\prime$~\cite{Larin:2010kq}, this possibility could become realistic. For instance, for $m_a = m_{\eta^{\prime}}$ and 
$g_{a\gamma\gamma} = 10^{-2}\,{\rm GeV}^{-1}$, we find that $0.1 \lesssim \Gamma(a\to\gamma\gamma)/\Gamma(\eta^{\prime}\to\gamma\gamma)\lesssim 0.2$. It has to be seen whether 
such effect is within JLab's resolutions. 
Instead, for $0.1\lesssim m_a({\rm GeV}) \lesssim 0.2$, we find that Primakoff type experiments already set the constraint $g_{a\gamma\gamma} \lesssim 0.005\,{\rm GeV}^{-1}$. 
For $m_a \lesssim 0.2 \,{\rm GeV}$, an even higher sensitivity to $g_{a\gamma\gamma}$ could be potentially reached at the PADME experiment in Frascati~\cite{Raggi:2014zpa}.

%%%%%%%%%%%%%%%%%%%%%%%%%%%%%%%%%%%%%%%%%%
\subsection{Conclusions}
%%%%%%%%%%%%%%%%%%%%%%%%%%%%%%%%%%%%%%%%%%

In this work, we have examined the contributions of ALPs to lepton dipole moments, both $g$-2 and EDMs.  We concentrated on the ALP mass range $\sim 0.1\!-\!1\,{\rm GeV}$, 
a region where the relatively loose 
constraints on ALP couplings to photons and leptons leave open the possibility of significant effects. Light-by-light pseudoscalar ALP loop effects were shown to resolve the muon $g$-2 discrepancy 
for ALP-$\gamma\gamma$ couplings near the published bound~\cite{Jaeckel:2015jla} of $\mathcal{O}(10^{-2} {\rm GeV^{-1}})$; but their effect drops quadratically with decreasing values, becoming negligible near $\mathcal{O}(10^{-3} {\rm GeV^{-1}})$. That is to be contrasted with Barr-Zee effective loop calculations where  the product of $g_{a\gamma\gamma} \,y_{a\mu} \sim 10^{-7}{\rm GeV^{-1}}$ provides a fairly robust solution to the muon $g$-2 discrepancy for a range of $g_{a\gamma\gamma}$ values extending down to $10^{-4} {\rm GeV^{-1}}$ 
(see fig.~\ref{fig:ma_cagg_plane}). 
Such large $g_{a\gamma\gamma}$ couplings are currently allowed by direct published~\cite{Jaeckel:2015jla} experimental constraints and perturbative unitarity. However, they can be better tested by new experiment at $e^+e^-$ facilities such as 
KLOE2~\cite{kloe2}, CMD3~\cite{cmd3}, SND~\cite{snd}, BES III~\cite{bes3} and Belle II~\cite{belle2} through dedicated searches for $e^+e^-\to e^+e^- a$ (see fig.~\ref{fig:sigma_gm2}).
In addition, a thorough analysis~\cite{Knapen:2016moh} of high energy $e^+e^- \to \gamma \gamma$ in LEPII data, including experimental acceptances and efficiencies, is likely to provide improved sensitivity to $g_{a\gamma\gamma}$ via $e^+e^- \to a \gamma \to 3 \gamma$. For a scalar ALP, the leading LbL contribution was found to have the wrong sign relative to the muon $g$-2 discrepancy. However, the BZ contribution could have either sign depending on the relative sign of $g_{s\gamma\gamma}$ and $y_{s\mu}$.  If only the LbL piece is considered (e.g. $y_{s\mu}$ effects assumed negligible), one can obtain the rather stringent indirect bound $g_{s\gamma\gamma} < \mathcal{O}(10^{-3}{\rm GeV}^{-1})$.
For CP violating ALP couplings, the electron EDM was found to probe much smaller, theoretically better accommodated ALP interactions over a range of parameters that overlap with the SHiP 
proposal~\cite{Alekhin:2015byh}. 
Future improvements in electron and nucleon EDM searches is strongly warranted both as a way to explore ALP parameters and to potentially unveil a new source of CP violation. Indeed, any new source of CP violation beyond SM expectations could impact our understanding of the matter-antimatter asymmetry of our Universe.    
     
%\end{document}

%%%%%%%%%%%%%%%%%%%%%%%%%%%%%%%%%%%%%%%%%%
\vspace{1cm}{\bf Acknowledgements}
%%%%%%%%%%%%%%%%%%%%%%%%%%%%%%%%%%%%%%%%%
We would like to thank G.~Abbiendi, S.~Eidelman, F.~Piccinini, M.~Raggi and A.~Wulzer, for very useful discussions. 
The research of A.M.\ and P.P.\ is supported by the ERC Advanced Grant No.\ 267985 (DaMeSyFla) and by the INFN. 
A.M. gratefully acknowledges support by the research grant Theoretical Astroparticle Physics No.  2012CPPPYP7 under the program PRIN 2012 funded by the MIUR.  
M.P.\ and P.P.\ acknowledge partial support by FP10 ITN Elusives (H2020-MSCA-ITN-2015-674896) and Invisibles-Plus 
(H2020-MSCA-RISE-2015-690575). This work is supported in part by the U.S.\ Department of Energy under Grant de-sc0012704.
M.P. is grateful to Columbia University for the hospitality during a visit when this manuscript was finalized.

%%%%%%%%%%%%%%%%%%%%%%%%%%%%%%%%%%%%%%%%%%

\end{document}